\documentclass[aps,prb,preprint,linenumbers,superscriptaddress, showpacs,showkeys,floatfix]{revtex4-1}
\usepackage{graphicx}
\usepackage{amsmath}
\usepackage{bm}
\usepackage{color}
\usepackage[version=3]{mhchem} 
\usepackage{chemfig}
\usepackage{array}

\begin{document}

\title{Growth and Electronic Structure of Boron-Doped Graphene}

\author{J. Gebhardt}
\email{these authors have contributed equally to this work}
\affiliation{Lehrstuhl f\"ur Theoretische Chemie, Universit\"at Erlangen-N\"urnberg, 91058 Erlangen, Germany}

\author{R. J. Koch}
\email{these authors have contributed equally to this work}
\affiliation{Lehrstuhl f\"ur Technische Physik, Universit\"at Erlangen-N\"urnberg, 91058 Erlangen, Germany}

\author{W. Zhao}
\affiliation{Lehrstuhl f\"ur Physikalische Chemie II, Universit\"at Erlangen-N\"urnberg, 91058 Erlangen, Germany}

\author{O. H\"ofert}
\affiliation{Lehrstuhl f\"ur Physikalische Chemie II, Universit\"at Erlangen-N\"urnberg, 91058 Erlangen, Germany}

\author{K. Gotterbarm}
\affiliation{Lehrstuhl f\"ur Physikalische Chemie II, Universit\"at Erlangen-N\"urnberg, 91058 Erlangen, Germany}

\author{S. Mammadov}
\affiliation{Lehrstuhl f\"ur Technische Physik, Universit\"at Erlangen-N\"urnberg, 91058 Erlangen, Germany}

\author{C. Papp}
\email{christian.papp@chemie.uni-erlangen.de}
\affiliation{Lehrstuhl f\"ur Physikalische Chemie II, Universit\"at Erlangen-N\"urnberg, 91058 Erlangen, Germany}

\author{A. G\"orling}
\affiliation{Lehrstuhl f\"ur Theoretische Chemie, Universit\"at Erlangen-N\"urnberg, 91058 Erlangen, Germany}

\author{H.-P. Steinr\"uck}
\affiliation{Lehrstuhl f\"ur Physikalische Chemie II, Universit\"at Erlangen-N\"urnberg, 91058 Erlangen, Germany}

\author{Th. Seyller}
\affiliation{Lehrstuhl f\"ur Technische Physik, Universit\"at Erlangen-N\"urnberg, 91058 Erlangen, Germany}

\date{\today}

\def\DEGC{$^\circ$C}
\def\angst{\,\text{\AA}}
\newcommand{\tief}[1]{$_{\text{#1}}$}
\newcommand{\hoch}[1]{$^{\text{#1}}$}

\newcommand{\correct}[1]{\textbf{!!!} \textcolor{red}{#1} \textbf{!!!}}

\begin{abstract}
The doping of graphene to tune its electronic structure is essential for its further use in carbon based electronics. Adapting strategies from classical silicon based semiconductor technology, we use the incorporation of heteroatoms in the 2D graphene network as a straightforward way to achieve this goal. Here, we report on the synthesis of boron-doped graphene on Ni(111) in a chemical vapor deposition process of triethylborane on the one hand and by segregation of boron from the bulk on the other hand. The chemical environment of boron was determined by x-ray photoelectron spectroscopy and angle resolved photoelectron spectroscopy was used to analyze the impact on the band structure. Doping with boron leads to a shift of the graphene bands to lower binding energies. The shift depends on the doping concentration and for a doping level of 0.3~ML a shift of up to 1.2~eV is observed. The experimental results are in agreement with density-functional calculations. Furthermore, our calculations suggest that doping with boron leads to graphene preferentially adsorbed in the top-fcc geometry, since the boron atoms in the graphene lattice are then adsorbed at substrate fcc-hollow sites. The smaller adsorption distance of boron compared to carbon leads to a bending of the graphene sheet in the vicinity of the boron atoms. By comparing calculations of doped and undoped graphene on Ni(111), as well as the respective free-standing cases, we are able to distinguish between the effects that doping and adsorption have on the band structure of graphene. Both, doping and bonding to the surface, result in opposing shifts on the graphene bands.
\end{abstract}

\maketitle

\section{Introduction}
The concept of doping was initially introduced in classical semiconductor technology, but was also adapted for carbon allotropes \cite{Zhou1994, Carroll1998, Endo2001, Hishiyama2001, Duclaux2002}. Doping of carbon material with nitrogen and boron atoms generated intense interest due to the possibility of tailoring the physical properties, i.e., electronic and transport properties \cite{Carroll1998, Endo2001, Hishiyama2001, Martins2007} but also chemical properties like the ability to adsorb lithium for capacitors \cite{Endo1998} or hydrogen storage capabilities \cite{Firlej2009}. In particular, these concepts were discussed recently for the two dimensional carbon allotrope graphene. The doping of graphene with nitrogen was achieved from nitrogen doped precursors or from post growth treatment procedures \cite{Wang2009, Wang2010, Zhao2011a, Koch2012, Usachov2011, Zhao2012}. The resulting new material showed promising first results towards applications in the field of electrochemical sensing \cite{Wang2009, Ci2010, Palnitkar2010, Marconcini2012}, lithium batteries \cite{
Reddy2010}, material in $p$-$n$ junctions \cite{Cheianov2007}, and fuel cells \cite{Qu2010}. Nitrogen doping of graphene showed a fundamental dependence on the geometry/site of the dopant. Besides the expected $n$ type behavior for substitutional doped graphene, $p$ type doping was observed for a dopant geometry, where nitrogen is introduced next to a carbon vacancy, forming pyridine-like units within the graphene lattice \cite{Koch2012, Schiros2012}. The changes in the band structure upon nitrogen-doping showed, nevertheless, the possibility of tuning the graphene band structure. The thermal stability of the resulting nitrogen-modified graphene sheets is similar to that of graphene.
Based on these results the incorporation of boron was also considered and first results from boron-doped graphene were already presented \cite{Tang2012, Kim2012}. In both casese exfoliated graphene was used, implying the already known challenges for the quality of the graphene layers and for large scale production. Also first results from the incorporation boron in grapheneoxide are reported. \cite{Sheng2012}

Herein, we report the production of single layer boron-doped graphene on a Ni(111) surface by chemical vapor deposition (CVD) of the boron-containing precursor triethylborane (TEB) and from segregation. The two processes used yield tunable concentrations of boron in a highly ordered graphene layer. The boron-doped graphene is characterized with angle resolved photoemission spectroscopy (ARPES) and x-ray photoelectron spectroscopy (XPS). The experimental results are analyzed and discussed by comparison with density-functional theory (DFT) calculations of undoped an boron-doped graphene on either Ni(111) or free-standing in vacuum, respectively. Our data show a strong shift of the graphene bands of up to 1.2~eV with respect to the Fermi level, depending on the boron concentration. During the doping process the overall band structure is retained. The comparison to the DFT calculations shows excellent agreement with the experiments.  Our calculations also suggest that doping with boron leads to graphene being adsorbed in the top-fcc geometry, because the boron atoms in the graphene lattice preferentially adsorb in the substrate fcc-hollow sites. The smaller adsorption distance of boron atoms within graphene compared to carbon atoms leads to a bending of the graphene sheet in the vicinity of the boron atoms. By comparing free-standing and/or undoped graphene with the adsorbed boron-doped graphene we are able to discuss the effects that doping and adsorption have on the band structure separately. Both modifications result in opposing shifts on the graphene bands. For high dopant concentrations this results in a Dirac point above the Fermi level.

\section{Experimental Setup and Computational Details}

The experiments were conducted at the third generation synchrotron source BESSY II in Berlin, Germany. Boron-doped graphene on Ni(111) was produced by CVD using TEB as precursor at temperatures between 600 and 950~K, leading to boron concentrations of typically 0.15 up to 0.35~ML after an exposure of 1800~L. Concentrations below 0.15~ML were prepared by segregating boron from the bulk, while exposing the nickel crystal to propene at 10\hoch{-6}~mbar at 900~K until saturation of the carbon signal. The boron is dissolved in the bulk by exposure to TEB and subsequent annealing to temperatures of 1100~K. The ARPES measurements were carried out at beamline U 56/2 PGM 2 using a Phoibos 100 analyzer  \cite{Koch2012}, while a transportable set up was used for the XPS measurements at beamline U~49/2 PGM~1 \cite{Denecke2002}. All XP spectra are taken with a photon energy of 380~eV and an overall resolution of 200~meV at normal emission. The boron concentration was calibrated from the adsorption of TEB at 130~K: we compared the TEB carbon intensity to the known intensity of a saturated benzene layer on Ni(111) at 200 K \cite{Papp2006}. From this we calibrated the boron coverage from the known boron content in TEB. The ARPES data was recorded using a photon energy of 70~eV. The DFT calculations were carried out using the VASP program package \cite{Kresse1996}, employing a plane wave basis and the Projector-Augmented-Wave (PAW) method for treating the core electrons \cite{Blochl1994}. The Perdew-Burke-Ernzerhof (PBE) \cite{Perdew1996} functional, including a correction to take dispersive forces into account \cite{Ortmann2006}, was applied. This theoretical setup was shown to accurately describe the 
interaction of graphene with metal surfaces\cite{Zhao2011,Kozlov2012,Mittendorfer2011}. The graphene/metal systems were modeled by six layer slabs of nickel. During optimization, the topmost three nickel layers were relaxed to model the surface, while the bottom three layers were fixed at the calculated nickel bulk positions. The unit cell of the adsorbed graphene containing two carbon atoms was, in accordance with experiments, chosen to be commensurable with the nickel surface unit cell containing one nickel atom. Different boron coverages were simulated by replacing one carbon atom by a boron atom in the graphene sheet of (2$\times$2), (3$\times$3), and (4$\times$4) graphene unit cells, leading to dopant concentrations with respect to the nickel atoms in the first substrate layer of 0.25, 0.11, and 0.063~ML, respectively. In the (2$\times$2) cells $13\times13\times1$ Monkhorst-Pack\cite{Monkhorst1976} \textbf{k}-point grids and respective smaller ones for the larger unit cells were used. They ensured a convergence with respect to total energies and band energies within $\pm0.01$~eV. For further details on the theoretical setup see \cite{Koch2012}.

\section{Results and discussion}
\subsection{Preparation, Doping Geometry, and Doping Level}

\begin{figure}[ht]
 \includegraphics[width=0.46\textwidth]{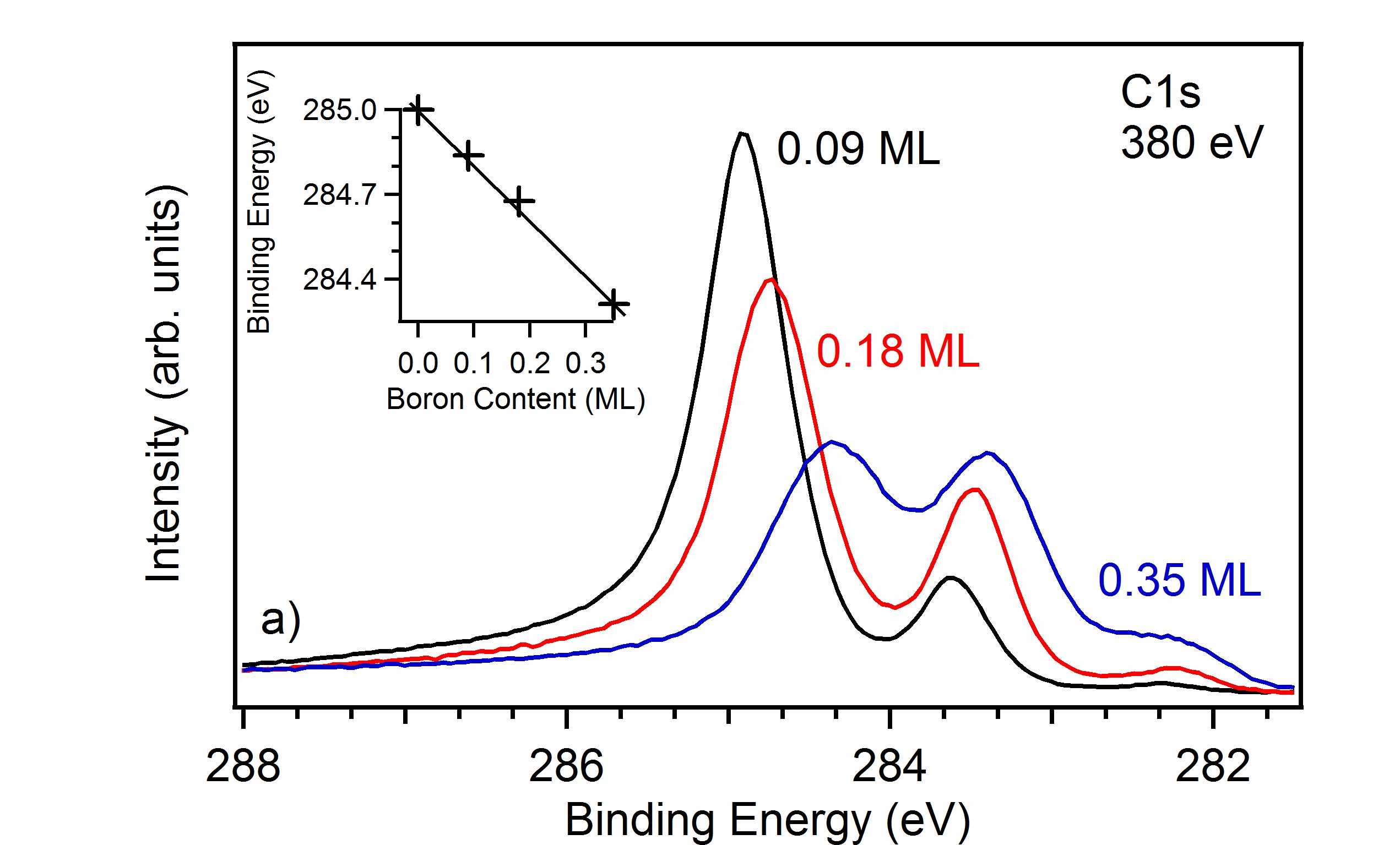}
 \includegraphics[width=0.46\textwidth]{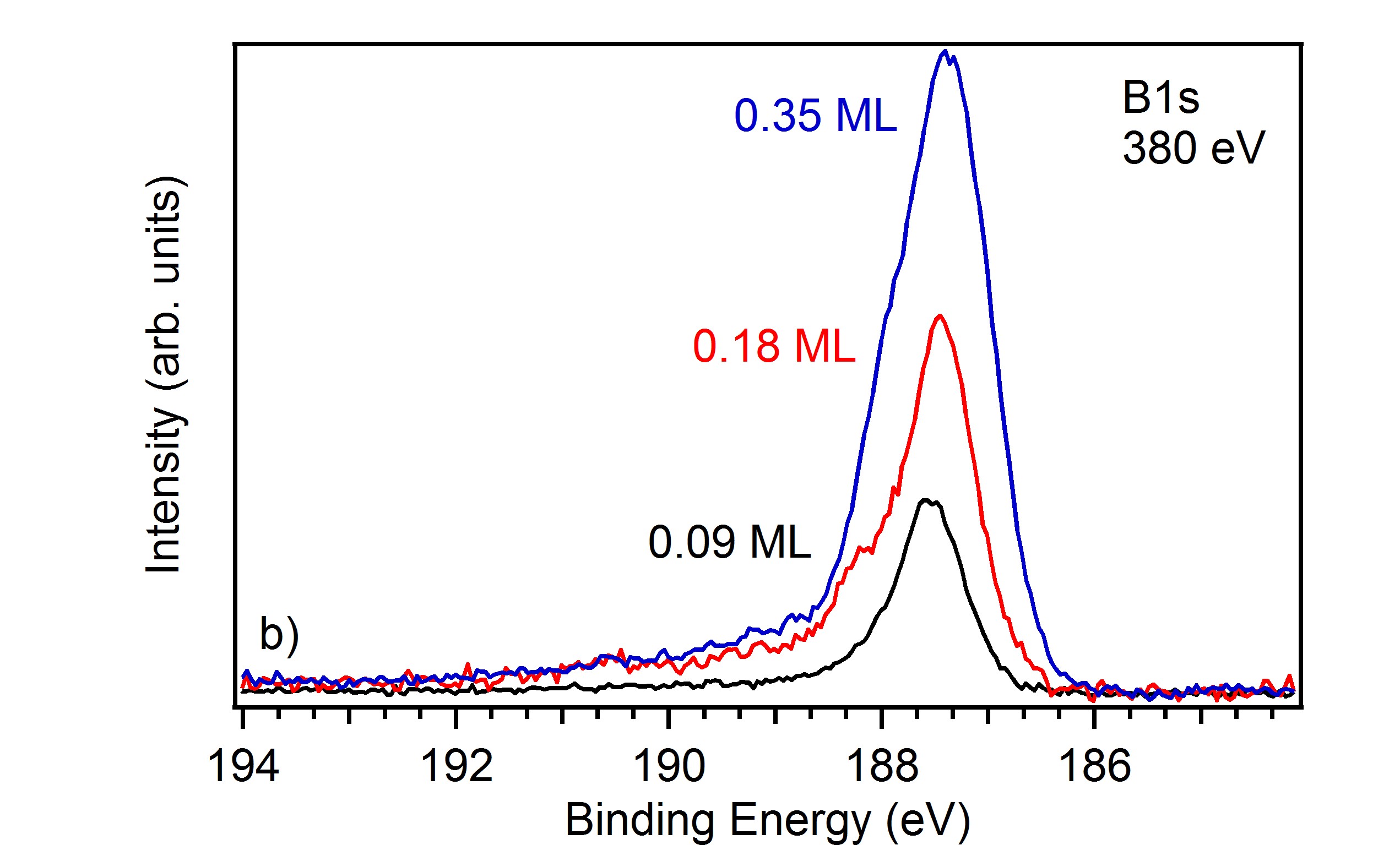}
 \caption{High resolution XP spectra of (a) the C $1s$ and (b) the B $1s$ region of boron-doped graphene with different doping levels.  The inset in (a) shows the shift of the graphene C $1s$ peak with the doping concentration.}
 \label{XPS}
\end{figure}

Figs.~\ref{XPS}~(a) and (b) show typical C $1s$ and B $1s$ XP spectra of boron-doped graphene layers grown on Ni(111). In the C $1s$ region two main peaks are observed, the graphene peak at $\mathtt{\sim}$285.0~eV and a second one shifted to lower binding energies by about $\mathtt{\sim}$1.6~eV. For further analysis, we calculated core level shifts (CLS) of the C $1s$ and the B $1s$ states according to the final state approximation\cite{Kohler2004}, for different possible arrangements (Tab.~\ref{CLS}). From comparison of the calculated and the experimentally observed shifts in the C $1s$ spectra the peak at lower binding energies is attributed to the formation of C\tief{2}CB, i.e., substitutional boron (theoretical shift of 1.42~eV). 
The shift towards lower binding energies is caused by the increased charge density on the carbon atoms due to their larger electronegativity in comparison to the adjacent boron atom. This is also in line with a Bader analysis\cite{Henkelman2006} carried out for the calculated systems that yields a charge transfer of 1.8 $e$ from each boron atom to the neighboring carbon atoms. The formation of a new peak in the C $1s$ region is in line with the results obtained for nitrogen doped graphene, where a new signal was observed at about 0.7~eV higher binding energies compared to the bridge-top graphene peak \cite{Zhao2012}. These shifts are expected considering  the electronegativities of carbon, nitrogen, and boron. 
Besides the formation of this additional peak in the C $1s$ spectrum, due to the bonds of carbon to the electropositive boron atoms, we additionally observe a shift of the main peak from 285 to 284.4~eV with rising boron coverage (see inset of Fig.~\ref{XPS}~(a)). The reason for this is that the boron dopants are influencing not only the C~$1s$ level of direct neighbors. A shift is also observed in the CLS analysis of carbon atoms in next neighbor spheres (Tab.~\ref{CLS}). This means that boron has also an influence on C\tief{3}C carbon atoms resulting in a shift and a broadening of the C $1s$ peak with increasing doping concentration. At even higher boron coverages a third peak at 282.2~eV is rising. The assignment of this smaller contribution is not unambiguously, but according to our CLS calculation we assign it to carbon bound to two substituted boron atoms (shift of 3.16~eV in Tab.~\ref{CLS}) Therefore, we suggest that primarily substitutional boron-doped graphene is formed, which is in line with the typical coordination sphere of boron, the applied growth mechanism, and is confirmed by our DFT calculations discussed in the following.

In the B $1s$ region there are two main contributions, located at 187.4~eV and a shoulder at  188.0~eV, respectively. They are attributed to boron bound to carbon, i.e., the species C\tief{2}CB, and elemental boron, respectively.
Please note that according to the calculated CLS the binding energy  of the B $1s$ electrons of boron  bound to a different amount of carbon atoms (C\tief{2}CB, CCB\tief{2}, and CB\tief{3}) are very similar. This makes a clear assignment of the boron geometry from the B $1s$ CLS difficult and, therefore, such conclusions were drawn from the data of the C $1s$ region. From a quantitative analysis of the C $1s$ and B $1s$ data the combined carbon and boron coverage on the surface was calculated to be 2 ML, as expected for a closed monolayer of graphene on Ni(111).

\begin{table}[ht]
\centering
	\begin{tabular}{>{\centering\arraybackslash}m{.14\linewidth}>{\centering\arraybackslash}m{.18\linewidth}|>{\centering\arraybackslash}m{.3\linewidth}>{\centering\arraybackslash}m{.3\linewidth}}
	Species&&  C $1s$ (eV) &  B $1s$ (eV) \\
        \hline
	C\tief{2}CB:&\chemfig{C(-[:0,-0.5])(-[:-60,0.5])-[:60,0.5]C(-[:120,0.5]C(-[:0,-0.5]-[:60,-0.5]-[:-60,0.5])-[:60,0.5]-[:0,0.5]-[:-60,0.5])-[:0,0.5]B(-[:60,0.5])-[:-60,0.5]-[:60,-0.5]-[:0,-0.5]}\vspace{0.1cm} & 1.42 (0.95) & -0.81 \\

	CCB\tief{2}:&\chemfig{B(-[:0,-0.5])(-[:-60,0.5])-[:60,0.5]C(-[:120,0.5]C(-[:0,-0.5]-[:60,-0.5]-[:-60,0.5])-[:60,0.5]-[:0,0.5]-[:-60,0.5])-[:0,0.5]B(-[:60,0.5])-[:-60,0.5]-[:60,-0.5]-[:0,-0.5]}\vspace{0.1cm} & 3.16 & -0.06 \\

	CB\tief{3}:&\chemfig{B(-[:0,-0.5])(-[:-60,0.5])-[:60,0.5]C(-[:120,0.5]B(-[:0,-0.5]-[:60,-0.5]-[:-60,0.5])-[:60,0.5]-[:0,0.5]-[:-60,0.5])-[:0,0.5]B(-[:60,0.5])-[:-60,0.5]-[:60,-0.5]-[:0,-0.5]}\vspace{0.1cm}& 4.17 & 0.41 \\

	C\tief{2}C:&\chemfig{(-[:0,-0.5])(-[:-60,0.5]-[:0,0.5]-[:60,0.5])-[:60,0.5]C(-[:120,0.5](-[:0,-0.5]-[:60,-0.5]-[:-60,0.5])-[:60,0.5]-[:0,0.5]-[:-60,0.5])}\vspace{0.1cm}& 2.86 &  ---\\
	\text{C\tief{2}CB-B:}&\chemfig{C(-[:0,-0.5])(-[:-60,0.5])-[:60,0.5]C(-[:120,0.5]C(-[:0,-0.5]-[:60,-0.5]-[:-60,0.5])-[:60,0.5]-[:0,0.5]-[:-60,0.5]B)-[:0,0.5]B(-[:60,0.5]B)-[:-60,0.5]-[:60,-0.5]-[:0,-0.5]}& 2.04 & 0.53 \\
	\end{tabular}
\caption {Calculated C $1s$ (relative to graphene in top-fcc adsorption) and B $1s$ (relative to a single boron atom) core level shifts for different geometrical arrangements within the (2$\times$2) cell of graphene adsorbed top-fcc on Ni(111). Different arrangements in the graphene network of a carbon atom with one, two, and three boron neighbors (C\tief{2}CB, CCB\tief{2}, and CB\tief{3}), a carbon atom next to a one atom vacancy (C\tief{2}C), and a carbon atom adjacent to a boron-boron bond (C\tief{2}CB-B) were calculated. The sketches display the relevant neighbors within the graphene lattice. The calculated C $1s$ shift always refers to the central carbon atom in the shown sketch (the value in brakets in the C\tief{2}CB case refers to the shift of the carbon atoms bond to the central carbon, i.e., carbon atoms C\tief{3}C with three neighboring carbon atoms). A positive shift refers to a shift towards lower binding energies. }
\label{CLS}
\end{table}

\subsection{Influence of the Boron Doping on the Geometry}
Since graphene is known to adsorb on Ni(111) in bridge-top and top-fcc adsorption geometries with almost identical binding energies\cite{Zhao2011, Fuentes-Cabrera2008}, both arrangements were considered in the calculations of graphene and substitutional boron-doped graphene in all unit cells. During geometry optimization, an interesting shift of the adsorption geometry was observed. While for undoped graphene both adsorption geometries turned out to be local minima on the potential energy surface, the top-fcc geometry is found to be the most stable arrangement in all doped cases. The geometry optimizations of substitutional boron-doped graphene, which were started in bridge-top geometry, converged to the top-fcc geometry. The energy gain of top-fcc compared to bridge-top adsorption is estimated to be 0.3-0.4~eV per unit cell in all cases, by comparing the optimized structures to calculations without geometry optimization in the respective fixed bridge-top adsorption geometries. This is due to the fact that boron atoms prefer to be located over a fcc-hollow site, 
leading to the top-fcc arrangement for the whole graphene sheet, although the carbon atoms are known to slightly prefer the bridge-top over the top-fcc arrangement \cite{Zhao2011}. Note, however, that the adsorption geometry could not be determined in our experiments, since the graphene C $1s$ and the broad C\tief{2}CB C $1s$ peak are superimposed and the latter dominates the splitting of the graphene C $1s$ peak in the case of the top-fcc geometry. Nevertheless, this geometry is found most likely in our preparations, since the boron is not introduced after graphene growth, but is present during growth. In addition, the observed shift from bridge-top towards top-fcc adsorbed graphene layers might also be possible in other preparation techniques, starting from primarily bridge-top adsorbed graphene, since the energetic difference between bridge-top and top-fcc adsorbed graphene is known to be small\cite{Zhao2011}. The graphene band structure, however, is known to be rather insensitive to the two possible geometric arrangements, neglecting fine differences in the close vicinity of the Dirac point and thus the observed shifts are equal for both geometries. This is confirmed by carrying out band structure calculations also of the unrelaxed bridge-top structures. The observed band shift of the optimized top-fcc structures is qualitatively as well as quantitatively reproduced. The differences of $<$0.06~eV are negligible and are due to the smaller distance of the graphene layer to the substrate in the case of the optimized top-fcc calculations.
An additional effect observed during geometry optimization is that the boron atom is, in the doped cases, adsorbed closer to the surface than carbon atoms in undoped graphene that are adsorbed in 2.12~$\angst$ from the nickel substrate. In the (2$\times$2) cell (0.25~ML) the adsorption distance of the boron atom is 2.02~$\angst$, i.e., the distance to the substrate is decreased by 0.1~$\angst$. The doping by boron atoms also affects the carbon atoms, which, in the (2$\times$2) cell, exhibit a decreased adsorption distance of $\mathtt{\sim}2.09 \angst$. In both larger unit cells the boron atom is adsorbed even closer to the nickel surface (1.99~$\angst$). This leads to a bending of the graphene layer: In the (4$\times$4) cell the carbon atoms are adsorbed at distances between $2.08$ and $2.14 \angst$, i.e., the carbon atoms that are directly bound to the boron atom are adsorbed closer to the surface, while the carbon atoms that are located further away from the boron atom are adsorbed at slightly increased adsorption distances, compared to undoped graphene.

\subsection{Influence of Doping and Adsorption on the Graphene Band Structure}

It is known that the energetic position of graphene bands on metal surfaces depends crucially on the adsorption distance\cite{Giovannetti2008, Khomyakov2009, Gong2010, Gebhardt2012}, i.e., the strength of the adsorbate-substrate interactions. Therefore, the effects on the band structure due to the above discussed small change in the adsorption distance in the case of doping and due to the incorporation of boron into the graphene geometry, were studied. To that end we compared the boron-doped graphene layer in the (2$\times$2) cell with and without taking into account the geometry changes due to boron-doping in the two adsorption distances of the optimized substitutional graphene (2.08~$\angst$) and the optimized undoped graphene (2.12~$\angst$). Both effects, the change in the adsorption distance and the geometry changes within the graphene sheet, influence the graphene band structure only negligible $<0.1$~eV. This means, we can exclude the possibility that the observed shifting of carbon bands is originating from geometrical changes due to the boron incorporation. Instead, this shift has to originate from the different chemical and physical properties of the boron dopants compared to the carbon atoms in the graphene network, in particular the effect of the dopants on the electronic structure. 

In the following, we first discuss the effect of boron-doping as observed in our experiments. Secondly, based on the DFT band structure calculations, we analyze in details, on the one hand, the impact of doping free-standing or adsorbed graphene and, on the other hand, of the effect of adsorbing pristine or doped graphene on the Ni(111) substrate.

\begin{figure}[ht]
 \includegraphics[width=0.46\textwidth]{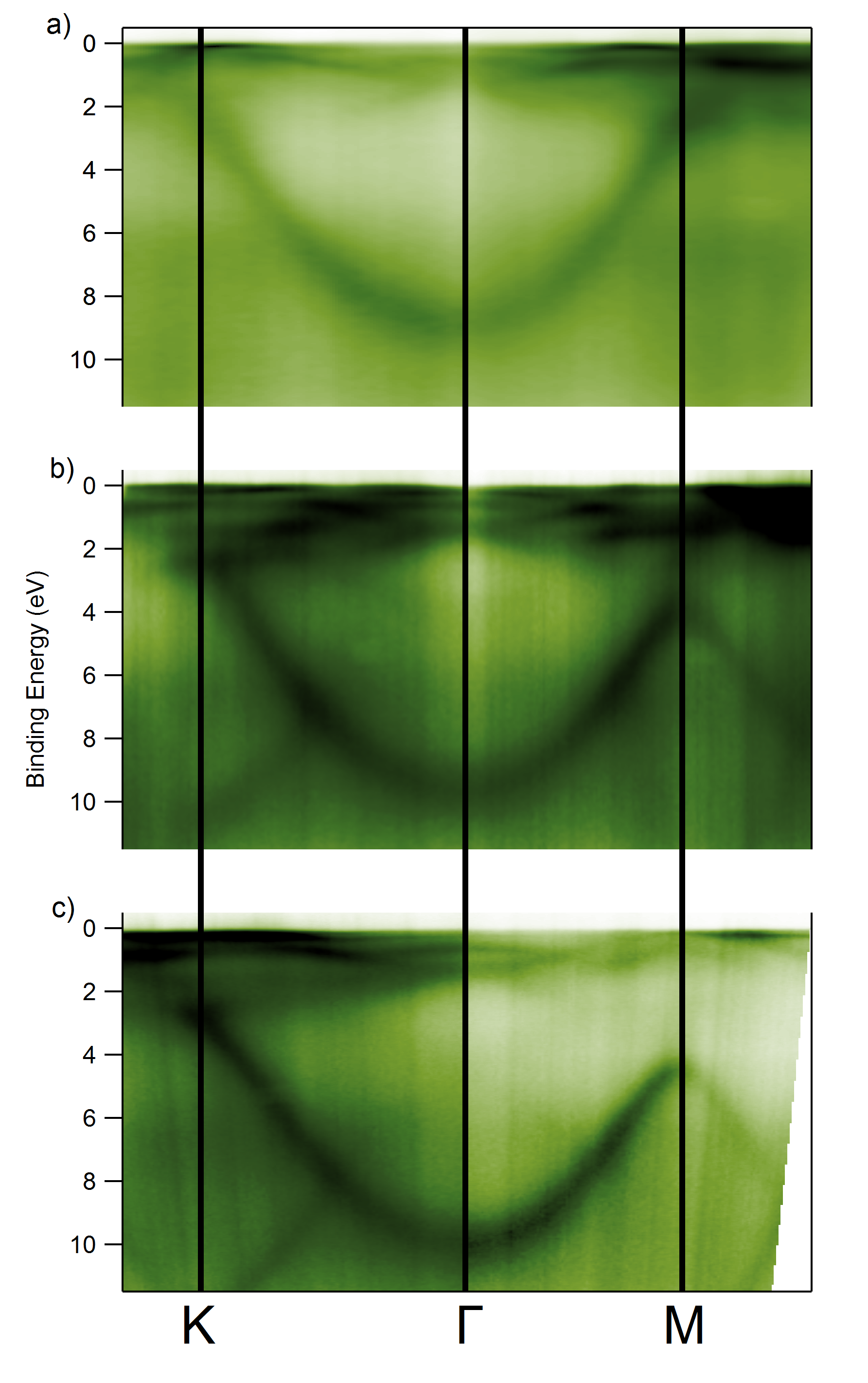}
 \caption{ARPES measurement of boron doped graphene with boron concentrations of (a) 0.28~ML, (b) 0.09~ML, and (c) 0.045~ML. }
 \label{ARPES} 
\end{figure}

\begin{figure}[ht]
\includegraphics[width=0.46\textwidth]{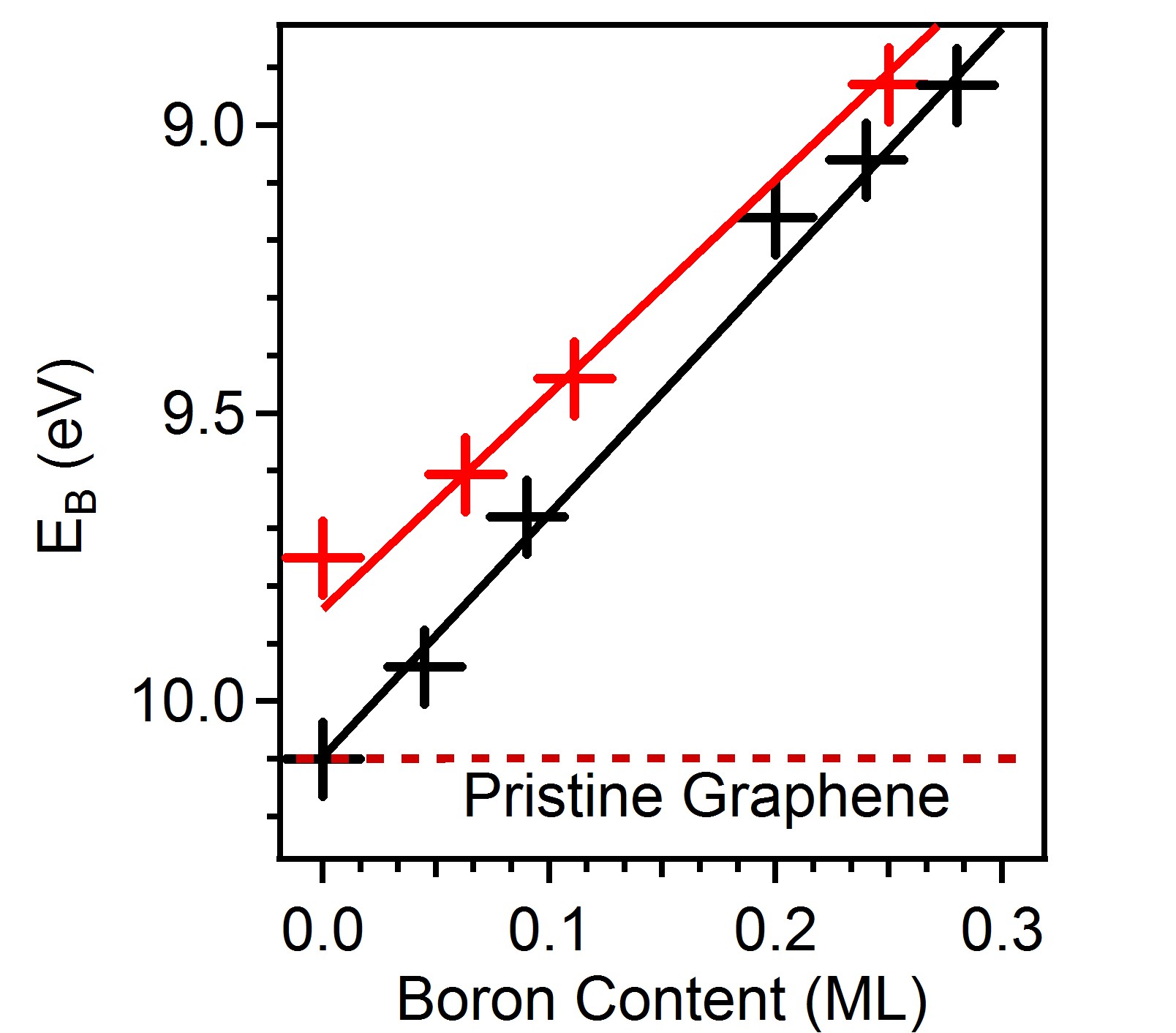}
\caption{\label{COMP} Shift of the binding energy E\tief{B} with the boron content of the carbon $\pi$ band at the $\bm \Gamma$ point of graphene on Ni(111) as measured in experiment (black) and predicted by theory (red).}
\end{figure}

In Fig.~\ref{ARPES} the ARPES data of boron-doped graphene for different dopant concentrations are displayed. Fig.~\ref{ARPES}~(c) corresponds to a doping of 0.045 ML. The typical band dispersion of graphene on Ni(111) is observed\cite{Varykhalov2008, Bertoni2004, Kozlov2012, Koch2012, Gebhardt2012}, but the energetic position of the $\pi$ band is shifted by 0.16~eV to smaller binding energies, in comparison to undoped graphene on nickel. The nickel bands are also visible, especially in the  binding energy region between 0 and 2~eV, where they couple to graphene states of appropriate symmetry \cite{Bertoni2004}. The spectra in Figs.~\ref{ARPES}~(a) and (b), which show higher doping concentrations, reveal a shift of the bottom of the $\pi$ band to subsequently smaller binding energies, depending on the doping level, up to a maximal shift of 1.2~eV in the case of a doping with 0.29~MLs of boron (Fig.~\ref{ARPES}~(a)). A quantitative analysis of this shift is given in Fig.~\ref{COMP}. Interestingly, we find that the spectra at higher doping levels become more diffuse, due to the incorporation of boron. In the data this is observed as a lower contrast in the shifted band structure. 
This is attributed to the fact that boron atoms, besides from doping the layer,  are also defects leading to new states localized around the original graphene bands. Due to the shift of the graphene bands to lower binding energies, we also observe hybridization of the $\pi$ band at the $\mathbf{M}$ point with nickel $d$ bands. Note that we did not find a change in the band dispersion or similar effects in our analysis of the ARPES data, i.e., the shift to lower binding energies is the only observed effect.

\begin{figure}[ht]
 \includegraphics[scale=0.15]{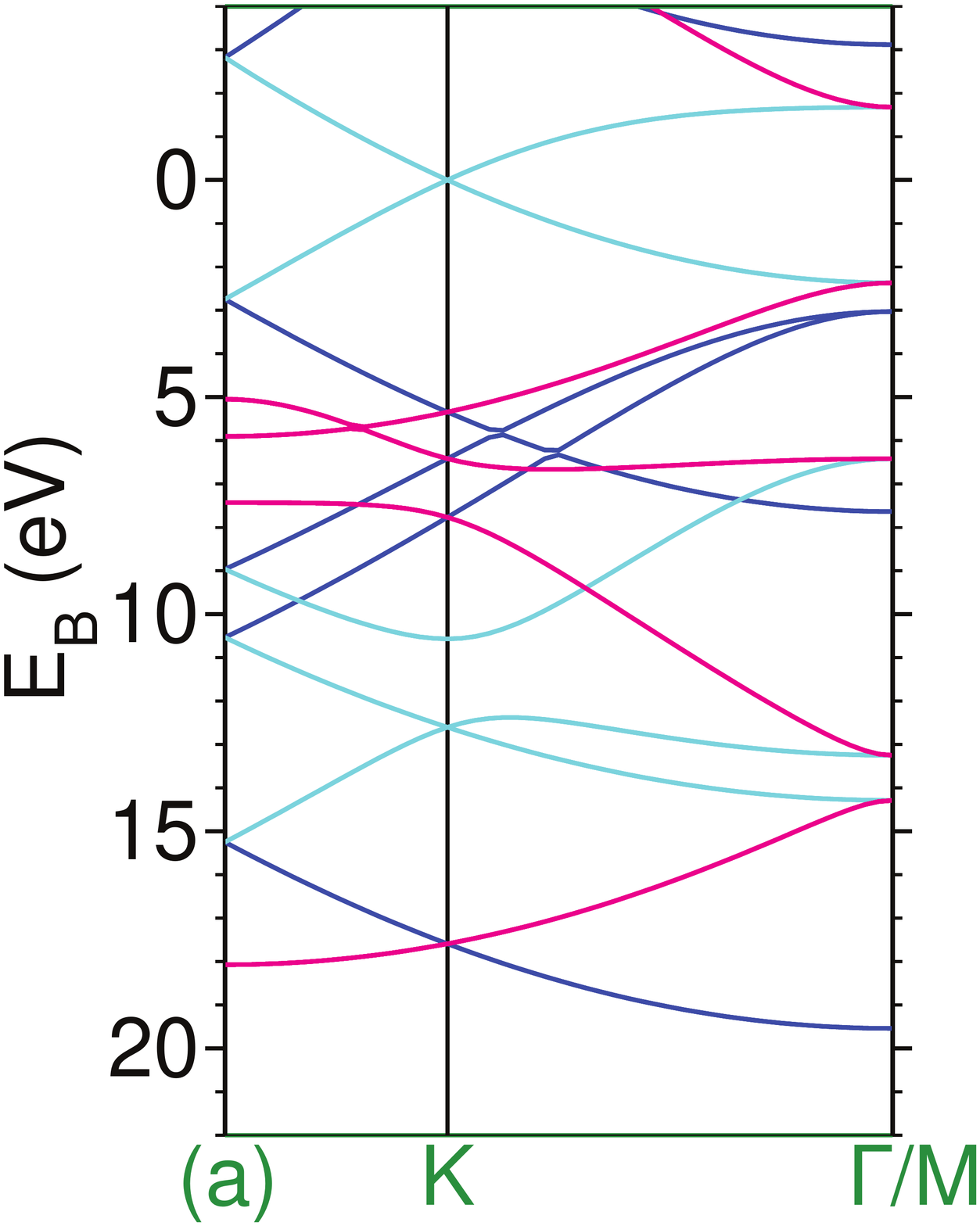}
\includegraphics[scale=1]{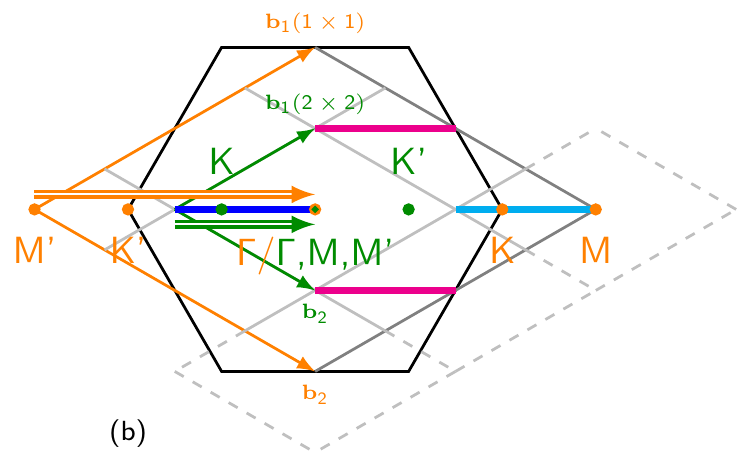}
 \includegraphics[origin=br,angle=-90,scale=0.15]{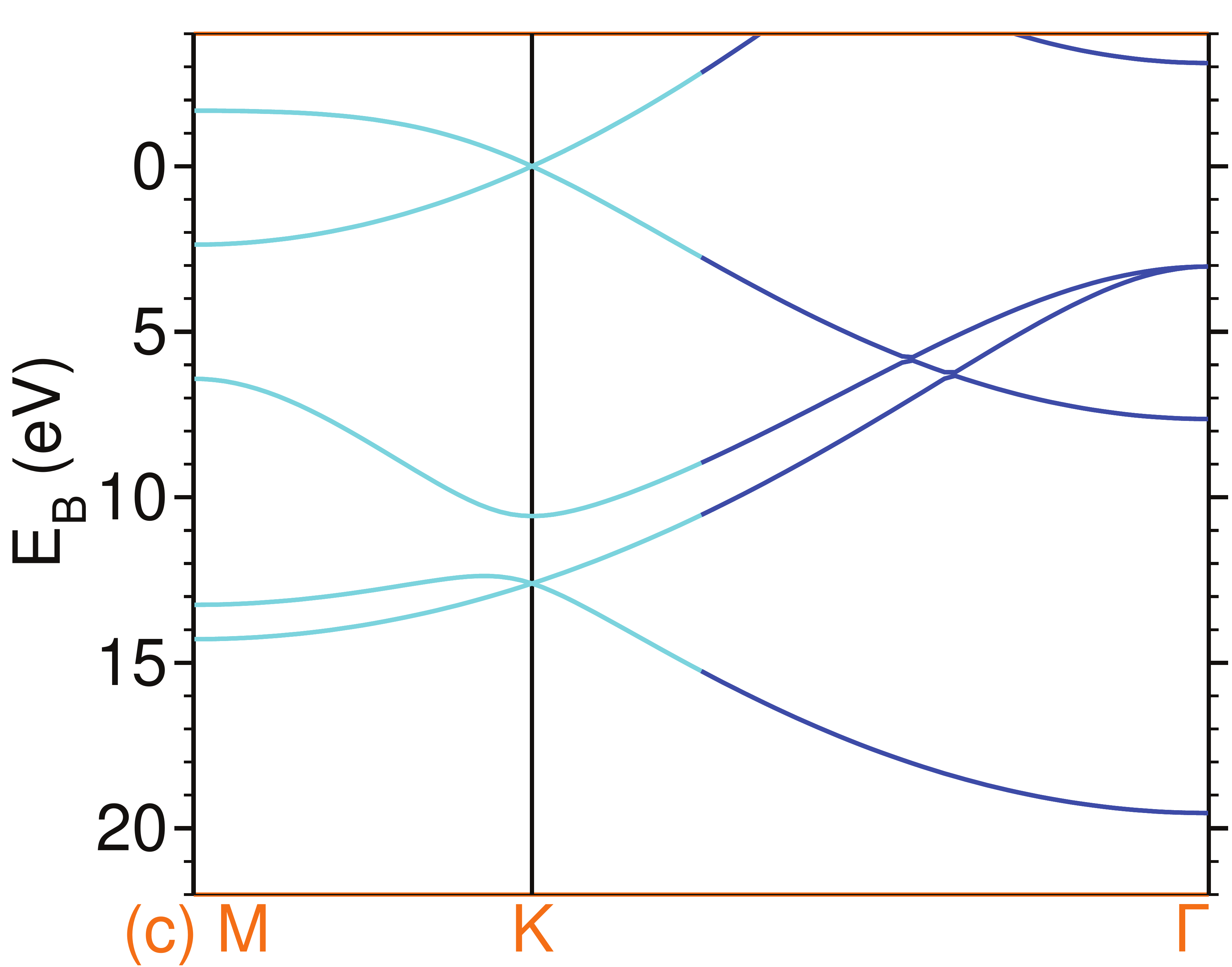}
\caption{The sketch in the middle displays the reciprocal cell to a (1$\times$1) hexagonal graphene unit cell (dark gray) within its Brillouin zone (black) and the respective reciprocal cell to a (2$\times$2) unit cell (light gray). The unit cell vectors and the positions of high symmetry points $\mathbf{M}$, $\mathbf{M}'$, $\mathbf{K}$, $\mathbf{K}$', and $\bm \Gamma$ are drawn orange and green for the reciprocal (1$\times$1) and (2$\times$2) unit cell, respectively. 
The bands along the paths colored in blue, cyan, and magenta in the reciprocal (1$\times$1) unit cell are backfolded on the path marked as green arrow in the reciprocal (2$\times$2) unit cell. To the left the band structure of free-standing, pristine graphene in the reciprocal (2$\times$2) unit cell along the green marked arrow is displayed with the color coding designating the corresponding paths in the reciprocal (1$\times$1) unit cell the bands are backfolded from. To the right the band structure of free-standing pristine graphene in the reciprocal (1$\times$1) unit cell along the orange arrow is displayed with the color coding relating the bands to the corresponding ones displayed in the reciprocal (2$\times$2) unit cell. (The cyan bands are the mirror image to those displayed in the reciprocal (2$\times$2) unit cell.)\label{BSTHEOEXPLAIN}}
\end{figure}

\begin{figure}[ht]
\includegraphics[width=0.22\textwidth]{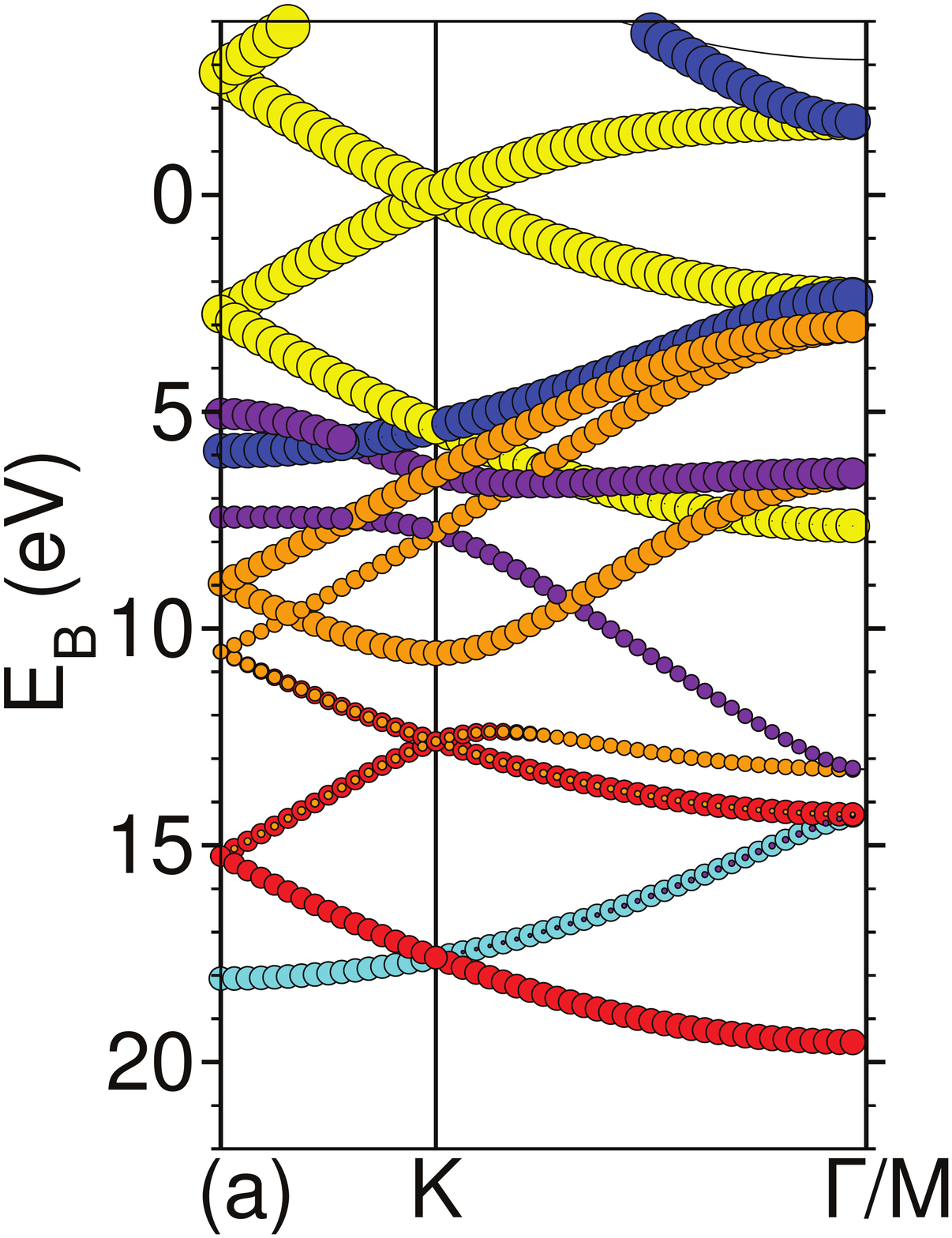}
\includegraphics[width=0.22\textwidth]{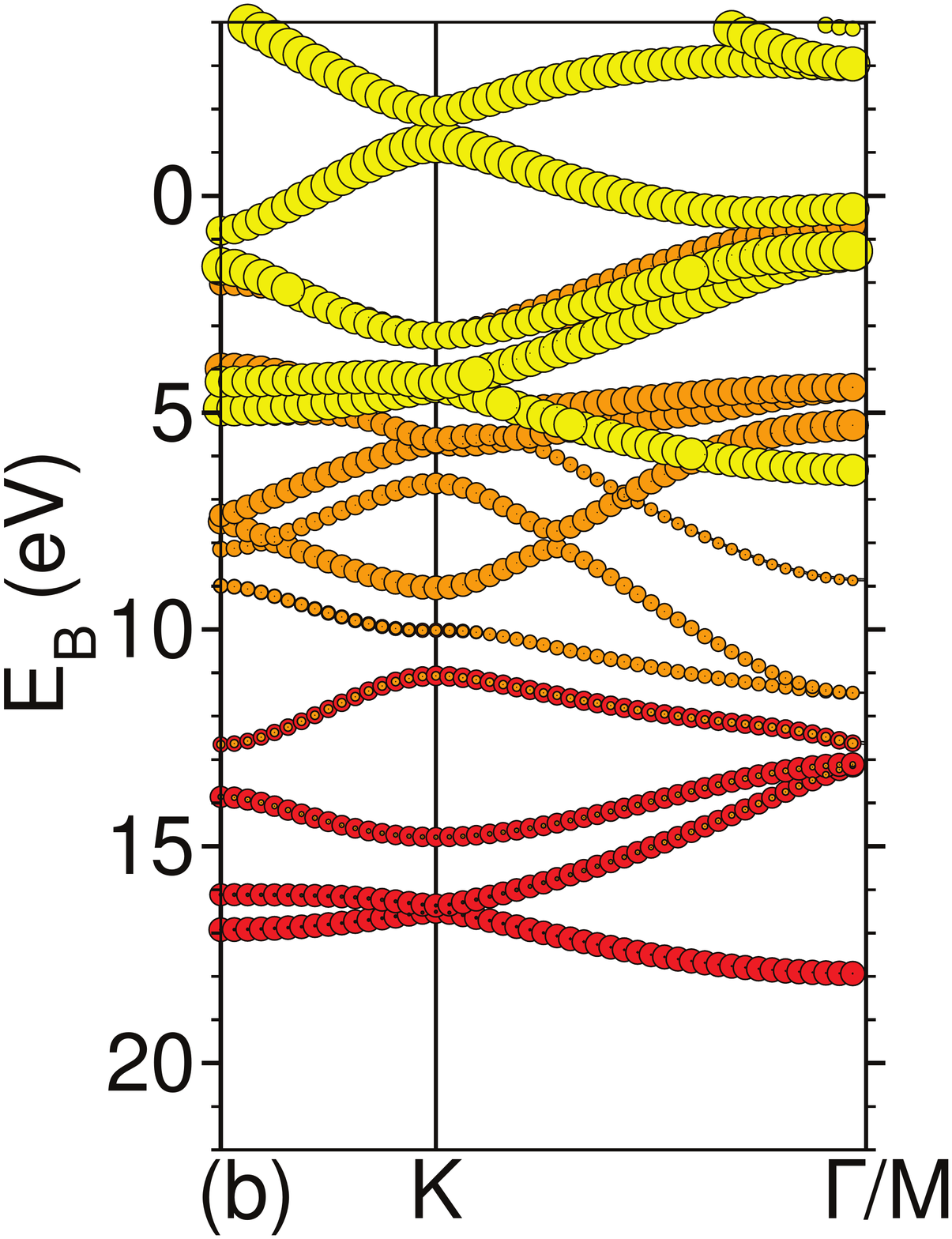}
\includegraphics[width=0.22\textwidth]{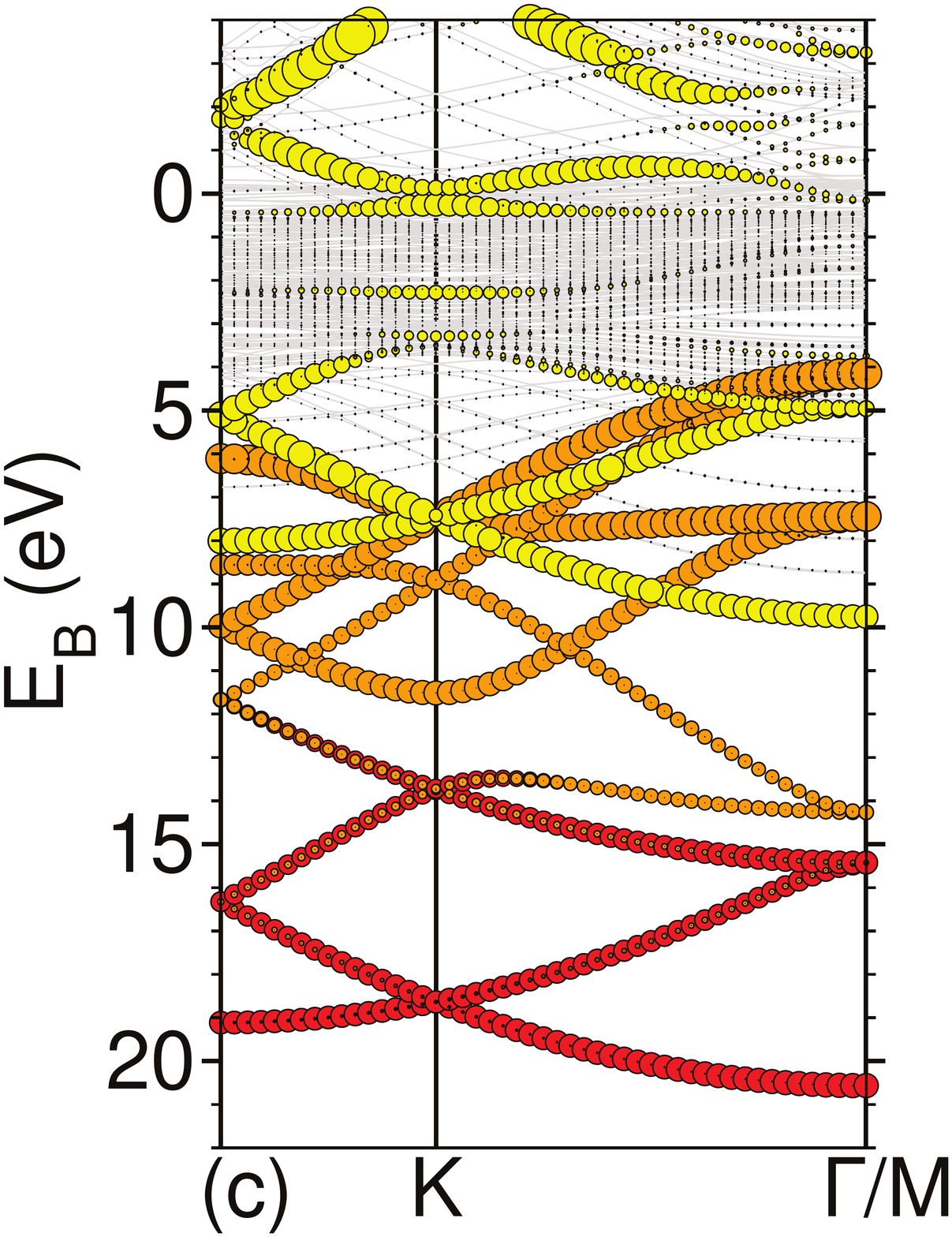}
\includegraphics[width=0.22\textwidth]{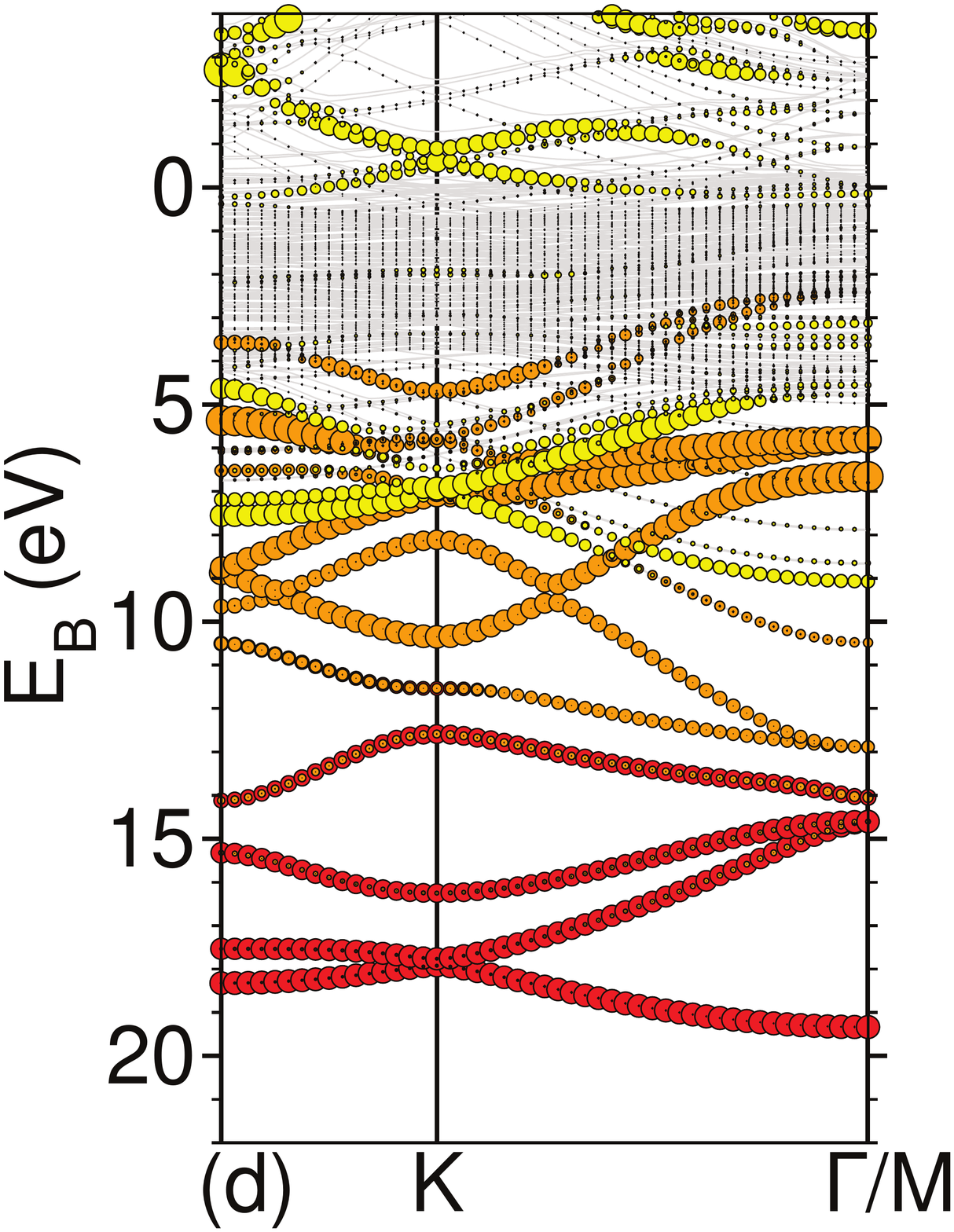}
\caption{\label{BSTHEO} Band structures of free-standing pristine graphene (a), free-standing graphene doped substitutionally by boron (b), pristine graphene adsorbed top-fcc on Ni(111) (c), and  boron-doped graphene adsorbed top-fcc on Ni(111) (d) calculated in (2$\times$2) unit cells. Red, orange, and yellow circles represent carbon 2$s$, 2$p_x$/2$p_y$, and 2$p_z$ contributions to carbon $s$, $\sigma$, and $\pi$ bands, with their radius being correlated to the manitude of the contribution to the bands (the radii of $p_z$ contributions are doubled for better visualization). Color coding is different for pristine graphene calculated in vacuum (a). Here, carbon 2$s$, 2$p_x$/2$p_y$, and 2$p_z$ contributions to bands that are backfolded from paths colored in Fig.~\ref{BSTHEOEXPLAIN}~(b) in magenta in the reciprocal cell corresponding to the (1$\times$1) unit cell are marked by cyan, purple, and blue circles, respectively. Nickel bands are represented by gray lines.}
\end{figure}

The experimental band structures in our study of boron-doped graphene adsorbed on Ni(111) are altered in two different ways compared to free-standing graphene. That is, on the one hand, the effects on graphene exerted by the nickel substrate and, on the other hand, the effects on graphene due to the dopant atoms.
In the following, these effects are discussed separately by considering the calculated band structures for the case of maximal doping, i.e., the (2$\times$2) unit cells, where the effects are most distinct. Four cases are considered, see Fig.~\ref{BSTHEO}: (a) free-standing, pristine graphene, (b) free-standing, boron-doped graphene, (c) pristine graphene adsorbed top-fcc on Ni(111), (d) boron-doped graphene adsorbed top-fcc on Ni(111). In Fig.~\ref{BSTHEO} band 
structures along the path marked in green in the reciprocal cell 
corresponding to the calculated (2$\times$2) unit cell are  shown, see
Fig.~\ref{BSTHEOEXPLAIN}~(b) for the different involved reciprocal cells and paths.

The paths marked in Fig.~\ref{BSTHEOEXPLAIN}~(b) in blue, cyan, and magenta, in the reciprocal cell corresponding to a (1$\times$1) graphene unit cell are  backfolded on the path marked as green arrow of the smaller (2$\times$2) reciprocal unit cell which underlies Fig.~\ref{BSTHEO}. Note, however, that the {\bf k} points denoted in the legend of Fig.~\ref{BSTHEO} refer to the original, large (1$\times$1) reciprocal cell of graphene. (The $\mathbf{\Gamma}$, {\bf M}, and {\bf M}' points of the reciprocal cell corresponding to the (1$\times$1) unit cell are backfolded on top of each other in the reciprocal cell corresponding to the (2$\times$2) unit cell.) For pristine graphene
the bands along the two magenta-marked paths are identical due to symmetry. In case of doping this symmetry is destroyed and the bands along these two paths become different. This shows up in additional bands in Fig.~\ref{BSTHEO} in the doped cases (b) and (d) compared to the undoped cases (a) and (c). The experimental band structure of Fig.~\ref{ARPES} is the one along the path marked as orange arrow in Fig.~\ref{BSTHEOEXPLAIN}~(b) in the (1$\times$1) reciprocal unit cell of graphene.

Comparison of the band structures of pristine and doped free-standing graphene,
Fig.~\ref{BSTHEO}~(a) and (b) shows that doping has two effects. Firstly, the shape of all bands is somewhat changed leading, among other things, to an opening of the Dirac point and, as discussed above, to the appearance of additional bands due to the lowering of the symmetry upon doping, i.e., upon replacing one carbon atom in the (2$\times$2) unit cell by boron.
Secondly, doping leads to a general shift of all graphene valence bands towards lower binding energies by about 1.2-2.0 eV with respect to the Fermi energy. This shift can be interpreted as classical doping effect. In the limit of classical doping the band structure is assumed to not change at all but the number of electrons in the system is modified. If we neglect the change of the bands in our case, then the replacement of carbon by boron decreases the number of electrons, i.e., leads to $p$ doping, which is accompanied by a lowering of the Fermi energy. The latter is tantamount to a shift of the graphene bands to lower binding energies with respect to the Fermi level, as found in the calculation and displayed in
Fig.~\ref{BSTHEO}. 

By comparing Fig.~\ref{BSTHEO}~(c) and (d) the effect of doping for the case of graphene adsorbed top-fcc on Ni(111) can by analyzed. It shows that the effect of  doping is very similar to the case of free-standing graphene. Again, the form of the graphene bands is somewhat changed and all graphene bands are shifted to lower binding energies. Now, however, this shift can not be simply explained by a lowering of the Fermi energy, because the latter is determined by the substrate bands, that is the nickel bands. In order to understand why doping with boron leads to a shift of the graphene bands to lower binding energies also in the case of graphene adsorbed on nickel, we disregard the change of the form of the bands for a moment. If we assumed that nickel and graphene bands, as well as their alignment, remains completely unchanged upon doping then the number of electrons on the graphene would not change because the Fermi level determined by the semi-infinite substrate would not change. This, however, would mean that the graphene layer would be negatively charged, by one electron per (2$\times$2) unit cell, because the replacement of boron by carbon reduces the positive charges of the nuclei of the graphene sheet by this magnitude. The electrons charging the graphene would come from the nickel substrate. This means the nickel surface would be charged by the same magnitude as the graphene sheet, which would lead to a dipole layer. The electrostatic potential of such a dipole layer can be easily calculated as the potential of a plate capacitor and in the considered case of a charge of one electron per (2$\times$2) unit cell would amount to a potential step of 17.5~eV between the nickel surface and the graphene layer (see, e.g., Ref.~\onlinecite{Gebhardt2012}). Raising of the graphene bands by 17.5~eV,
of course, is completely unphysical and, indeed would result in an almost complete depopulation of the graphene valence bands and in a positive changing much higher than the initial negative charging of the graphene layer and, furthermore, would be in contradiction to the assumption that the alignment between nickel and graphene bands remains unchanged upon doping of graphene with boron. In order to avoid such a contradiction we now allow this level alignment to change in the course of doping. Then, the following can happen if we still disregard changes of the form of bands upon doping: Only a small amount of charge is transferred from the nickel surface to the doped graphene sheet. The potential step of this dipole layer then shifts the graphene bands to lower binding energies similarly as in the case of free-standing graphene.
Indeed, shifts of the graphene bands of 1.75~eV, which is roughly the amount observed in free-standing graphene, would require only a transfer of 0.1 electron from nickel to graphene per (2$\times$2) unit cell. Actually, in the considered scenario, the shift of the graphene bands to lower binding energies with respect to the Fermi level upon doping has to be somewhat smaller in the case of graphene adsorbed on nickel than in the case of free-standing graphene because otherwise the graphene would be neutral again and no shift would occur. Therefore, one would expect a charging of the graphene layer upon doping by somewhat less than 0.1 electron per (2$\times$2) unit cell. This, indeed, is found in our calculations. The charge of the pristine graphene on Ni (111), according to a Bader charge analysis, is 0.352 electrons per (2$\times$2) unit cell, while that of the doped graphene on Ni(111) is 0.424 electrons per (2$\times$2) unit cell. This means doping leads to an increase of the negative charge of 0.072 electrons per (2$\times$2) unit cell.

In summary we can explain the shift of graphene bands to lower binding energies with respect to the Fermi level in the cases of free-standing and of graphene adsorbed top-fcc on Ni(111) as follows: In the case of free-standing graphene the Fermi energy is lowered because of the smaller number of electrons due to doping, in the case of graphene adsorbed top-fcc on Ni(111) a small charge transfer from the nickel substrate to the graphene sheet occurs which leads to dipole layers accompanied by a potential step between the nickel surface and the graphene sheet that is responsible for the shift of the graphene bands. This  means doping has a similar effect for free-standing graphene and graphene adsorbed on nickel but the explanation for the effect is different in the two cases.

Comparing Fig.~\ref{BSTHEO}~(a) and (c) and Fig.~\ref{BSTHEO}~(b) and (d) gives information about the effect of adsorbing pristine or doped graphene, respectively, top-fcc on nickel. In both cases a lowering of the graphene bands to stronger binding energies is found that is accompanied by $n$ doping of the graphene sheet. According to Ref.~\onlinecite{Gebhardt2012} (see also Refs.~\onlinecite{Giovannetti2008, Khomyakov2009, Gong2010}), the reason for this band shift and the $n$ doping is due to the surface dipole layer of the nickel substrate and a rearrangement of nickel surface charge known as pillow effect \cite{Bagus2002, Vazquez2007}.  

Considering the joint influence of adsorption and of boron doping together, Fig.~\ref{BSTHEO}~(a) \textit{vs.} (d), adsorption leads to a shift of the graphene bands to higher binding energies, whereas doping has a reverse effect. For the considered case of a doping of 0.25~ML these superimposed effects yield $\pi$ and $\pi^*$ bands that are located above the majority of the substrate bands around \textbf{K}. This results in a Dirac point that is estimated, as midpoint between $\pi$ and $\pi^*$, to be located 0.7~eV above the Fermi level and that is opened by 0.3~eV.

\subsection{Influence of the Boron Concentration}

In Figs.~\ref{ARPES}~(a)-(c) typical ARPES data of the preparation of boron-doped graphene on Ni(111), with boron contents ranging from 0.045 to 0.28 ML, are shown. The band structure of graphene, especially the $\pi$ band and its typical dispersion leading upward from the $\bm \Gamma$ point to the \textbf{M} and the \textbf{K} point is visible in all three cases.
The main difference between the three cases is the energetic position of the minimum of the $\pi$ band at the $\bm \Gamma$ point with respect to the Fermi level.
Fig.~\ref{COMP} shows that this position is measured to decrease from 10.1~eV for pristine graphene to 8.93~eV for maximally doped graphene (doping of 0.28~ML). 

The energy shift from the ARPES measurements is in good agreement with the shift observed in corresponding DFT band structure calculations (the underestimation of the absolute binding energies in the DFT band structures by about 0.2-0.4~eV is typically and reflects the inaccuracy of binding energies from DFT calculations). This confirms the agreement between our explanations derived from calculations and our experimental results, i.e., it confirms that the boron-doped graphene was predominantly synthesized in the substitutional doping arrangement. The observed shift of the $\pi$ bands towards smaller binding energies is in line with the findings for nitrogen doped graphene, where the $\pi$ band is shifted towards higher binding energies for the substitutional doping geometry. 
In the band structure calculations of free-standing graphene, we find a similar shift (not displayed) of 1.45~eV of the position of the $\pi$ band at the $\bm \Gamma$ point, with respect to the Fermi level, for the case of a boron coverage of 0.25 ML relative to pristine graphene. In the free-standing doped cases the energetic position of the Dirac point is estimated as midpoint between the graphene $\pi$  and $\pi^*$ band at \textbf{K}, due to the absence of nickel bands. The bandwidth is estimated as difference between $E_D$ and the energy of the $\pi$ band at $\bm \Gamma$, which reveals a small increase of the width of the $\pi$ band of 0.26~eV. This shows, that the observed changes of the $\pi$ band at $\bm \Gamma$ are due to a shift of all graphene bands relative to the Fermi level and should not be mistaken with a decreased band dispersion.

\section{Conclusion}

We have shown that boron-doped graphene can be grown on a Ni(111) surface in a CVD process using the boron-containing precursor triethylborane or by segregation from a boron rich nickel crystal. Doping by boron leads to a strong shift of the graphene valence bands to lower binding energies.
Additionally, the graphene $\pi$ band becomes more diffuse in the case of high boron-doping.
Our results show a facile way of doping graphene with boron accompanied by a tuning of the band energies, up to shifts of 1.2~eV. It is found that the effect of the adsorption of graphene on Ni(111), which leads to a shift of the graphene bands to higher binding energies, is counteracted by introducing boron, resulting in an opened Dirac point in the unoccupied states for high boron coverages. Furthermore, our DFT calculations showed that boron-doped graphene prefers to adsorb in the top-fcc geometry, due to the strong preference of boron to adsorb in fcc-hollow sites. Due to the smaller bonding distance of boron compared to carbon, we predict a bending of the graphene layer in the case of low boron concentrations.

{\bf Acknowledgements}
The authors gratefully acknowledge the funding of the BMBF through grant 05 ES3XBA15 and the German Research Council (DFG) which supports the Collaborative Research Center 953 and which supports, within the framework of its Excellence Initiative, the Cluster of Excellence 'Engineering of Advanced Materials' (www.eam.uni-erlangen.de) at the University of Erlangen-Nuremberg. We thank the BESSY staff, especially W. Mahler and B. Zada at UE56-2, for their support during beamtime. W. Z. thanks the China Scholarship Council for financing his Ph. D. grant.


%

\end{document}